\documentclass[english]{article}

\usepackage{geometry} 
\usepackage{graphics}
\usepackage{amsmath}
\usepackage{amssymb}
\usepackage{epstopdf}
\usepackage{babel}
\usepackage{inputenc}
\usepackage{epsfig}

\begin{document}
\begin{center}
{\Large {\bf Results of the Search for Strange Quark Matter and Q-balls with the SLIM Experiment}} 

\vspace{0.8cm}

\normalsize{\small{
S. Cecchini$^{1,2}$, M. Cozzi$^{1,3}$, D. Di Ferdinando$^{3}$, M. Errico$^{1,3}$, F. Fabbri$^{3}$, G. Giacomelli$^{1,3}$, R. Giacomelli$^{3}$,\\ M. Giorgini$^{1,3}$, A. Kumar$^{1,4}$, J. McDonald$^{5}$, G. Mandrioli$^{3}$, S. Manzoor$^{1,6}$, A. Margiotta$^{1,3}$, E. Medinaceli$^{1,7}$,\\ L. Patrizii$^{3}$, J. Pinfold$^{5}$, V. Popa$^{3,8}$, I.E. Qureshi$^{6}$, O. Saavedra$^{9,10}$, Z. Sahnoun$^{3,11}$, G. Sirri$^{3}$, M. Spurio$^{1,3}$,\\ V. Togo$^{3}$, C. Valieri$^{3}$, A. Velarde$^{7}$ and A. Zanini$^{10}$}
\par~\par

{\footnotesize{\it
(1) Dip. Fisica dell'Universit\'a di Bologna, 40127 Bologna, 
Italy \\  
(2) INAF/IASF Sez. Bologna, 40129 Bologna, Italy\\
(3) INFN Sez. Bologna, 40127 Bologna, Italy\\
(4) Physics Dept., Sant Longowal Institute of Eng. \& Tech., Longowal, 
148 106, India\\  
(5) Centre for Subatomic Research, Univ. of Alberta, Edmonton, 
Alberta T6G 2N4, Canada\\ 
(6) PD, PINSTECH, P.O. Nilore, and COMSATS-CIIT, No. 30, H-8/1, Islamabad, Pakistan\\ 
(7) Laboratorio de F\'{i}sica C\'{o}smica de Chacaltaya, UMSA, La Paz, Bolivia\\ 
(8) Institute for Space Sciences, 77125 Bucharest, Romania\\
(9) Dip. Fisica Sperimentale e Generale, Universit\'a di 
Torino, 10125 Torino, Italy\\ 
(10) INFN Sez. Torino, 10125 Torino, Italy\\
(11) Astrophysics Dept., CRAAG, BP 63 Bouzareah, 16340 Algiers, Algeria}}
}

\end{center}

\vspace{1cm}

\abstract{\small{
The SLIM experiment at the Chacaltaya high altitude laboratory was sensitive to nuclearites and Q-balls, which could be present in the cosmic radiation as possible Dark Matter components. It was sensitive also to strangelets, i.e. small lumps of Strange Quark Matter predicted at such altitudes by various phenomenological models. The analysis of 427 m$^{2}$ of Nuclear Track Detectors exposed for 4.22 years showed no candidate event. New upper limits on the flux of downgoing nuclearites and Q-balls at the 90\% C.L. were established. The null result also restricts models for strangelets propagation through the Earth atmosphere.}
} 

\section{Introduction}
\label{sec:intro}
\normalsize
Nuggets of Strange Quark Matter (SQM) composed of approximately the same number of up, down and strange quarks could be the true ground state of quantum chromodynamics~\cite{Witten}. SQM nuggets could be stable for all baryon numbers in the range between ordinary heavy nuclei and neutron stars ($A\sim10^{57}$). They may have been produced in the early Universe~\cite{Witten}, and could be present in the cosmic radiation as a possible component of the galactic cold dark matter (with typical velocities of $\sim$$10^{-3}$c)~\cite{DeRujula}. As the strange quark is massive compared to almost massless up and down quarks, surface tensions lead to the suppression of a few s quarks. Thus, SQM should have a relatively small positive electric charge compared to that of heavy nuclei~\cite{Farhi,Heiselberg,Madsen98}. In what follows, macroscopic quark nuggets, neutralised by captured electrons are called \emph{nuclearites}. Otherwise, and generally for small baryon numbers ($A<10^{6}$), assumed to be quasi totally ionized, they will be called \emph{strangelets}. In some cases, the term \emph{SQM} or \emph{SQM bags} will be used interchangeably for nuclearites and strangelets if the properties discussed apply to both. In ref.~\cite{Madsen98} it was shown that an even more stable state of nuclear matter could exist, the so called ``Colour-Flavour Locked'' (CFL) SQM, because of the occurrence of a Cooper-like pairing between quarks of different colour and flavour quantum numbers. Strangelets are unlikely to have survived from the early Universe but they could be produced in the core of neutron stars and released into the Galaxy in very energetic astrophysical processes involving strange star collisions~\cite{Madsen88,Friedman} and/or supernovae explosions~\cite{Vucettich}.

	Other particles discussed in this paper are \emph{Q-balls}~\cite{Coleman}. They are hypothetical supersymmetric coherent states of squarks, sleptons and Higgs fields, predicted by minimal supersymmetric generalisations of the Standard Model of particle physics. \emph{Q-balls} could have been copiously produced in the early Universe and may have survived till now as a dark matter component. Their general properties are different from those of SQM but in some cases, as discussed later, the flux limits derived for nuclearites can apply also to them.
	
	The SLIM (Search for LIght Magnetic monopoles) experiment, operated for more than four years at the high altitude Chacaltaya laboratory in Bolivia (5230 m a.s.l.), was designed to search for light and intermediate mass magnetic monopoles~\cite{Cecchini05}. It was sensitive also to SQM and Q-balls~\cite{Balestra05} offering the possibility to extend the searches for these particles to a lower mass range.
	
	In the following, we present a short description of the apparatus and of the experimental procedure used for the analysis. The properties and energy losses of SQM and Q-balls in Nuclear Track Detectors (NTDs) are also discussed. Finally, we give the flux upper limits set by SLIM.

\section{Experimental}
\label{sec:experimental}
The SLIM experiment was a large array ($\sim$430 m$^{2}$ area) of Nuclear Track Detectors (NTDs) deployed at the Chacaltaya laboratory. The detector modules consisted of stacks composed of three layers of CR39$^{\scriptsize\textregistered}$, three layers of Makrofol DE$^{\scriptsize\textregistered}$, two layers of Lexan and a 1 mm thick aluminium absorber to slow down or stop nuclear recoils. The detectors were exposed to the cosmic radiation for an average of 4.22 years~\cite{Cecchini05} after which they were brought back to the Bologna Laboratory where they were etched in ``strong'' and ``soft'' etching conditions, as described in~\cite{Balestra05,Balestra07,Balestra0801}, and analyzed. 

	The strong etching conditions for CR39 were: 8N KOH + 1.5\% ethyl alcohol at 75~$^{\circ}$C for 30 hours. The soft etching conditions were : 6N NaOH + 1\% ethyl alcohol at 70~$^{\circ}$C for 40 hours. For Makrofol the etching conditions were 6N KOH + 20\% ethyl alcohol at 50~${^\circ}$C for 10 hours.

	We recall that the formation of etch-pit cones (``tracks'') in NTDs is regulated by the bulk etching rate, $v_{B}$, and the track etching rate, $v_{T}$, i.e. the velocities at which the undamaged and damaged materials (along the particle trajectory), are etched out. Etch-pit cones are formed if $v_{T} > v_{B}$. The response of the CR39 detectors is given by the etching rate ratio $p = v_{T}/v_{B}$.
	
	The CR39 and Makrofol nuclear track detectors were calibrated with 158 A GeV In$^{49+}$ and Pb$^{82+}$ beams at the CERN SPS and 1 A GeV Fe$^{26+}$ at the Brookhaven National Laboratory (BNL) Alternating Gradient Synchrotron (AGS). The calibration layout included a fragmentation target and CR39 (plus Makrofol) NTD foils in front of and behind the target~\cite{Cecchini08,Balestra0801}. The detector sheets behind the target detected both primary ions and nuclear fragments. After etching, the standard calibration procedure was the following: $(i)$ measurements of the base area of each beam and fragment tracks in CR39 with an automated image analyzer system~\cite{Elbek}; $(ii)$ for each fragment peak the $Z/\beta$  was obtained and the reduced etch rate ($p-1$) was computed. The Restricted Energy Loss (REL) due to ionization and nuclear scattering was evaluated, thus obtaining the calibration data expressed as ($p-1$) vs REL for all detected nuclear fragments~\cite{Balestra0801}.
	
	The threshold for CR39 was at $Z/\beta \sim 14$ ($\beta=v/c$) in strong etching conditions corresponding to REL $\sim 200$ MeV g$^{-1}$ cm$^{2}$. For soft etching the threshold was at $Z/\beta \sim 7$ which corresponds to REL $\sim 50$ MeV g$^{-1}$ cm$^{2}$.
	
	About 50 m$^{2}$ of the SLIM CR39 contained 0.1\% of DOP additive in order to reduce the neutron induced background by raising the detector threshold \cite{Tarle}. The threshold of CR39(DOP) was at $Z/\beta \sim 21$ in strong etching conditions corresponding to REL $\sim 460$ MeV g$^{-1}$ cm$^{2}$, and $Z/\beta \sim 13$ for soft etching, corresponding to REL $\sim 170$ MeV g$^{-1}$ cm$^{2}$. For  Makrofol the threshold was at $Z/\beta \sim 50$ (REL $\sim 2.5$ GeV g$^{-1}$ cm$^{2}$).
	
	The standard scanning procedure for the SLIM experiment was the following. After strong etching (applied to all uppermost CR39 layers of each stack) the CR39 foils were scanned twice by different operators, using a stereo microscope with a 3$\times$ magnification large optical lens. The signal looked for was a hole or a biconical track with the two base cone areas equal within experimental uncertainties. The double scan guarantees an efficiency of $\sim$100\% for finding a possible signal.
 The bottom CR39 sheets were softly etched when a track was considered as a possible ``candidate'' in the uppermost layers (This was about 10\% of the cases). The third CR39 sheet was etched only in few cases, when there was still a possible uncertainty. A few \% of Makrofol foils were etched for similar reasons. More details about the experimental analysis procedure of NTDs, are given in ref.~\cite{Balestra0801} and references therein.
 
	No candidate events remained after all cross checks. Two ``very unusual events'' were observed and were finally classified as manufacturing defects in a small subset of CR39 NTDs ($\sim1$m$^{2}$). These ``unusual events'' are discussed in detail in ref.~\cite{Balestra0802}.

\section{Energy Losses and Detection of Strange Quark Matter}
\label{sec:SQM}
In this section we shall recall the main properties of nuclearites and strangelets, their energy losses in NTDs and some propagation models in the Earth atmosphere. 

\subsection{Nuclearites}
\label{sub:Nuclearites}
Nuclearites are considered to be large strange quark nuggets, with overall neutrality ensured by an electron cloud which surrounds the nuclearite core, forming a sort of atom~\cite{DeRujula,Bakari}. Their radius is never smaller than $10^{-8}$ cm. As the core size increases the electron cloud is increasingly contained in the nuclearite core. If the nuclearite mass is larger than 1.5 ng ($8.4\cdot10^{14}~\textrm{GeV/c}^{2}$) all electrons are inside the core. 

\begin{figure*}
\begin{center}
\resizebox{!}{7.5cm}{\includegraphics{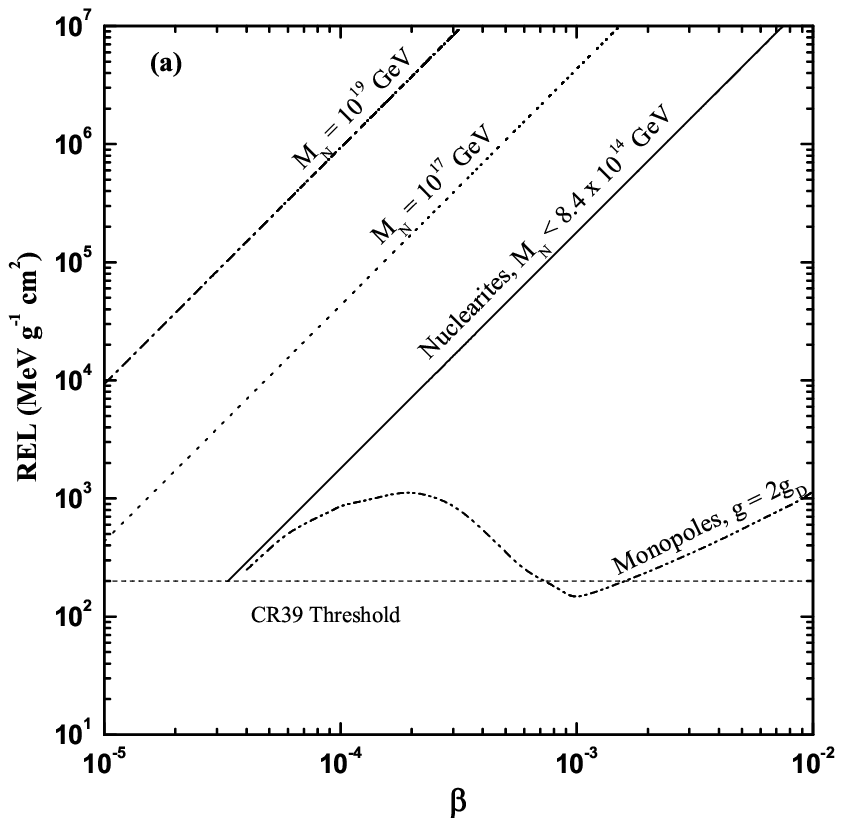}}
\hspace{0.5cm}
\resizebox{!}{7.5cm}{\includegraphics{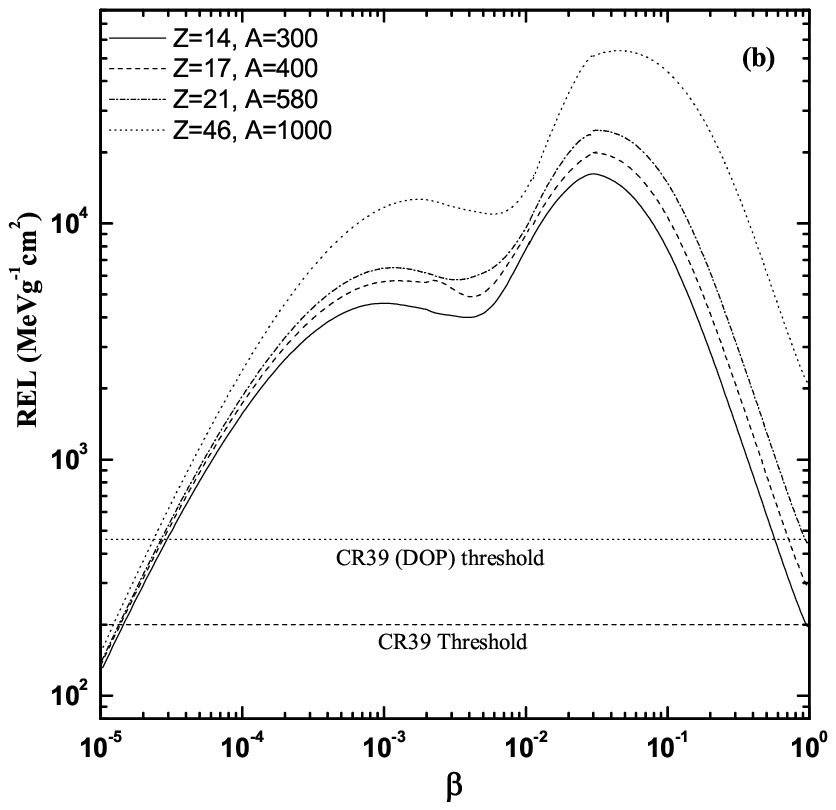}}
\caption{\small(a) REL versus $\beta$ for nuclearites of different masses. For comparison the REL of magnetic monopoles with charge $g=2g_{D}$ is also shown. (b) REL versus $\beta$ for strangelets with small electric charges and masses $(Z, A) = (14, 300), (17, 400), (21, 580)$ and $(46, 1000)$. The thresholds (strong etching) for CR39 and CR39(DOP) are indicated by the horizontal lines.}
\label{fig:REL}
\end{center}
\end{figure*}

In ref.~\cite{DeRujula} it was postulated that elastic or quasi elastic collisions with atoms and molecules of the traversed medium are the relevant energy losses of non-relativistic large mass nuclearites:
\begin{equation}
\label{eq:rel}
\frac{dE}{dx}= -\sigma \rho v^{2}
\end{equation}
where $\rho$ is the density of the traversed medium, $v$ the nuclearite velocity and $\sigma$ is its cross section:
							
\begin{equation}
\label{eq:sigma}
\sigma =\left\{ \begin{array}{ll}
\pi \cdot {10}^{-16} \,\textrm{cm$^{2}$}     & \qquad \text{for $ M < 1.5 \,$ ng}\\
 & \\
\pi {\left(\displaystyle{\frac{3 M}{4 \pi {\rho}_{N}} }\right)}^{2/3}     & \qquad \text{for $M > 1.5 \,$ ng}\
\end{array} \right.
\end{equation}
\mbox{$\rho_{N}= 3.5\cdot10^{14}\,g/cm^{3}$} is the nuclearite density and $M$ its mass.
	
 Fig.~\ref{fig:REL}a shows the REL of nuclearites in CR39 as a function of their initial velocity for different masses, computed using Eqs.~\ref{eq:rel}, \ref{eq:sigma}. It can be seen that the REL of any nuclearite with velocity $\beta > 3\cdot10^{-5}$  is above SLIM's detection threshold.
	
	A nuclearite with mass $M$ entering the atmosphere with an initial velocity $v_{0} << c$, after crossing a depth $L$, will be slowed down to

\begin{equation}
\label{eq:velocity}
v(L)=v_{0} \,\exp\left(-\frac{\sigma}{M} \int_{0}^{L} \rho \,dx \right)
\end{equation}
where $\rho$ is the air density at different depths and $\sigma$ the interaction cross section from Eq.~\ref{eq:sigma}.

\begin{figure}
\begin{center}
\resizebox{7.7cm}{7.7cm}{\includegraphics{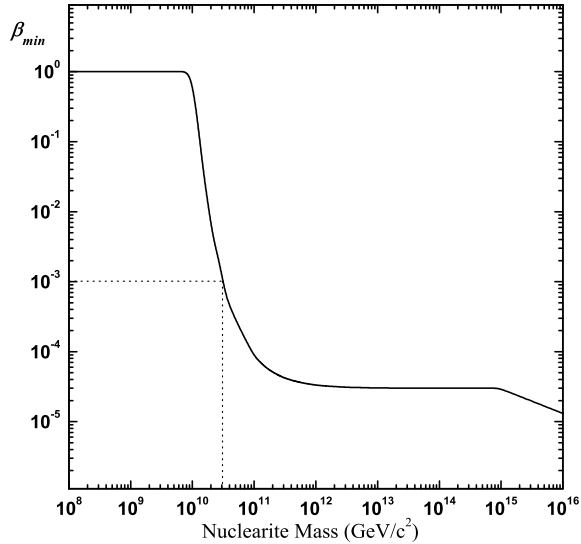}}
\caption{\small Minimum velocity for nuclearites at the top of the atmosphere to be detected by the SLIM experiment, versus their mass.  For $\beta_{min} \sim 10^{-3}$ all nuclearites with mass $\geq 3\cdot 10^{10}$ GeV/c$^{2}$ should be detected.}
\label{fig:acceptance}
\end{center}
\end{figure}

	The accessible region for the SLIM experiment at the Chacaltaya level (540 g/cm$^{2}$ of residual atmosphere) is expressed as the minimum incident velocity of nuclearites at the top of the atmosphere, which yields at the detector level a detectable signal (i.e. REL above the detector threshold) as can be seen in Fig.~\ref{fig:acceptance}. It is assumed, in the propagation through the Earth atmosphere, that the variation in mass of the nuclearite and gravitational effects, are negligible. For an initial velocity of $v_{0}\sim 10^{-3}c$, SLIM would have detected all nuclearites impinging on the detector with a mass equal or greater than $3\cdot 10^{10}$ GeV/$c^{2}$. 

\subsection{Strangelets}
\label{sub:Strangelets}
One of the most important properties of SQM bags is their charge to mass ratio, which is very small compared to ordinary nuclear matter. For ``ordinary'' strangelets, as opposed to CFL strangelets, the relation charge $Z$ to mass number $A$ is given by \cite{Heiselberg}:
\begin{equation}
\label{eq:mass}
\begin{array}{ll}
Z\approx 0.1~\left(\displaystyle{\frac{m_{s}}{150\,\text{MeV}}}\right)^{2}~A   & \qquad \text{for} \quad A\ll 10^{3} \\
  & \\
Z\approx 8~\left(\displaystyle{\frac{m_{s}}{150\,\text{MeV}}}\right)^{2}~A^{1/3}  & \qquad \text{for} \quad A\gg 10^{3}\
\end{array}
\end{equation}
while for Color-Flavor Locked strangelets it is~\cite{Madsen98} 

\begin{equation}
\label{eq:cflmass}
Z\approx 0.3~\left(\frac{m_{S}}{150\, \text{MeV}}\right)~A^{2/3}
\end{equation}
where $m_{s}$ is the strange quark mass usually considered to be $m_{s}\sim$$150$ MeV.
The $Z/A$ ratio is a key feature for the detection of strangelets.

	Strangelets should be stable for all baryon numbers exceeding a critical value $A_{crit}$$\sim$$300$~\cite{Farhi}. Although, it has been shown that some underlying shell effects may lead to smaller stable strangelets~\cite{Gilson}. They could be accelerated to relativistic velocities by the same astrophysical processes responsible for the acceleration of cosmic rays in the Galaxy. If we assume that strangelets have no associated electrons then their interaction with matter, in particular in NTDs, is expected to be similar to heavy ions \- but with a different $Z/A$. The REL of strangelets in CR39 nuclear track detectors is illustrated in Fig.~\ref{fig:REL}b. At low velocities we estimated it from ``Ziegler's fit'' to experimental data from low energy ion measurements~\cite{SRIM}.  At energies $>100$ keV/nucleon the REL was computed from~\cite{PDB}

\begin{align}
\label{eq:StrRel}
\textnormal{REL} &=  K Z^{2} \frac{Z_{t}}{A_{t}} \frac{1}{\beta^{2}} \times \left[ \frac{1}{2} \ln\left(\frac{2 m_{e}c^{2} \beta^{2} \gamma^{2} T_{upper}}{I^{2}}\right) \right. 
					\nonumber \\
                 &\quad  \left. - \frac{\beta^{2}}{2} \left(1+\frac{T_{upper}}{T_{max}}\right) - \frac{\delta}{2} - \frac{C_{s}}{Z_{t}} \right ]\
\end{align}

The terms in Eq.~\ref{eq:StrRel} are defined in ref.~\cite{PDB}.
$T_{upper} = \textnormal{MIN}(T_{cut}, T_{max})$ where for CR39 $T_{cut}$ = 200 eV and for Lexan and Makrofol $T_{cut}$ = 350 eV. 

	From Fig.~\ref{fig:REL}b, one can see that strangelets with charges above threshold and velocities in the range ($2\cdot10^{-5} \div 1$) $c$ can be detected with the SLIM CR39 detectors. However, since the threshold for CR39 (DOP) is higher, the velocity range is reduced to a maximum $\beta \sim0.7$ for charges lower than $Z\sim18$.

\begin{figure}
\begin{center}
\resizebox{7.7cm}{7.7cm}{\includegraphics{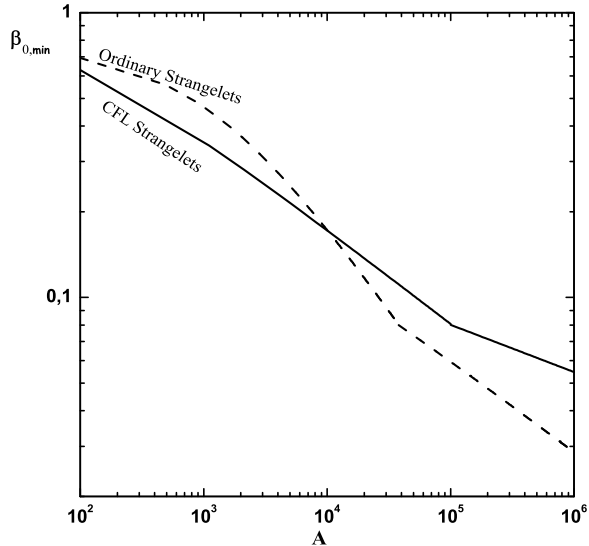}}
\caption{\small The minimum velocity strangelets should have at the top of the atmosphere to overcome the geomagnetic cutoff at Chacaltaya versus their mass. The solid line applies to CFL strangelets, the dashed line to ordinary ones. The change in slope (at $\sim 3\cdot10^{4}$ and $10^{5}$) is due to the solar modulation dominance over geomagnetic cut-off at large $A$.}
\label{fig:rigidity}
\end{center}
\end{figure}

	Being electrically charged, strangelets are affected by the geomagnetic field. In order to pass the geomagnetic cutoff at Chacaltaya ($R_{cutoff}$$\sim$$12.5$ GV~\cite{Smart}) strangelets must have a minimum velocity (at the top of the atmosphere) which depends on the baryon number, Fig.~\ref{fig:rigidity}. Note that for ordinary strangelets this minimum velocity was computed using the more accurate formula for the electric charge (see ref.~\cite{Heiselberg}). The above condition does not guarantee strangelet detection by ground based cosmic ray experiments as one has also to take into account their propagation through the Earth's atmosphere.

\subsubsection{Strangelet propagation models and detection modes for the detection at high altitude laboratories}
Two phenomenological models for the propagation of strangelets in the atmosphere were considered in this paper. 
	
	According to the model proposed by Wilk et al. \cite{Wilk}, large relativistic strangelets lose mass in each interaction with air nuclei as they travel through the Earth's atmosphere, until they reach their critical stability mass and begin to evaporate neutrons. The minimum mass at the top of the atmosphere for a strangelet to reach the Chacaltaya Laboratory was estimated to be $A_{0}$$\sim$$2200$ for normal incidence. The lightest strangelet at the detector level (with $A\sim$$300$) would have a charge of the order of $Z\sim14$ (Eqs.~\ref{eq:mass} and~\ref{eq:cflmass}). This is close to the SLIM threshold.  The flux estimated in ref.~\cite{WilkF} was of the order of $\sim$$2\cdot 10^{-13}$ cm$^{-2}$ s$^{-1}$ sr$^{-1}$ for initial strangelets of energies above 10 GeV. The expected number of events at the Chacaltaya laboratory would have been $\sim$ 6 strangelets per 100 m$^{2}$ per year for masses above $A_{crit}$ and energies above 1 GeV. This estimate was made assuming that the flux of strangelets is $F_{S}(A_{0}=A_{crit})\simeq2.4\cdot10^{-5} F_{tot}$ ($F_{tot}$ is the primary cosmic ray flux at the same energy per particle) and has an abundance spectrum, for $A>A_{crit}$, $\propto A$$^{-7.5}$.
	
	In the propagation model proposed by Banerjee et al.~\cite{Banerjee,BanerjeeF}, small stable or metastable strangelets (with $ A \sim 42, 54, 60, 84, 102, \ldots$) pick-up neutrons and protons as they traverse the Earth's atmosphere, increasing in charge and mass at different rates (proton absorption being quickly suppressed as the energy decreases). The strangelet would acquire its final charge at $\sim$$10$ km above sea level. It was estimated that an initial relativistic strangelet with charge $Z=2$ at the top of the atmosphere could attain a final charge of about $Z=14$. A larger initial charge ($>2$) at the top of the atmosphere could lead to larger final charges. The main requirement is that for this charge/mass acquisition process to occur the initial velocity of the strangelet should be lower than $0.7c$. Otherwise, the excitation energy of the strangelet, when colliding with atmospheric nuclei, could exceed the binding energy and lead to its fragmentation. Banerjee el al. gave an estimate of the flux of strangelets that could reach the level of Sandakphu in India (at $\sim3.6$ km a.s.l) for an experimental set-up equivalent to SLIM~\cite{BanerjeeF}. This flux, $5 \div 10$ events per 100 m$^{2}$ per year, could reasonably be extended at the level of Chacaltaya.
	
	Another phenomenological model for strangelet propagation in the atmosphere was proposed recently~\cite{Wu} but it was not considered in this work.
	
	In a recent paper, the flux above the rigidity cutoff for strangelets to be detected by on-board satellite instruments was estimated, assuming that they are produced in strange star collisions and then propagate through the Galaxy as ordinary cosmic ray particles \cite{Madsen05}. The expected flux is 
\begin{align}
\label{eq:flux}
F_{tot} &\sim 2\cdot10^{5} \,\text{m$^{-2}$yr$^{-1}$sr$^{-1}$} \, A^{-0.467} \, Z^{-1.2}
		\nonumber \\
        &\quad \times \text{max}[R_{SM},R_{GC}]^{-1.2} \Lambda ,\
\end{align}
where, $R_{GC}$ is the geomagnetic rigidity cutoff, $R_{SM}$ is the solar modulation cutoff and,

\begin{align}
\label{eq:lambda}
\Lambda &= \left(\frac{\beta_{SN}}{0.005}\right)^{1.2} \left(\frac{0.5\text{cm}^{-3}}{n}\right) \left(\frac{\dot{M}}{10^{-10}M_{\odot}\text{yr}^{-1}}\right)
		\nonumber \\
		& \quad  \left(\frac{1000\text{kpc}^{3}}{V}\right) \left(\frac{930\text{MeV}}{m_{0}c^{2}}\right). \
\end{align}
$\Lambda$ is highly dependent on the input parameters (the number of compact stars in binary systems, the average amount of mass released in strange star collisions, $\dot{M}$, compared to the solar mass $M_{\odot}$) which are poorly known. $\beta_{SN}$ is the speed of a typical supernova shock wave, $n$ the average hydrogen number density in the interstellar medium, $V$ the effective galactic volume and $m_{0}$ the nucleon mass. 

\section{Flux Upper Limits for SQM}
\label{sec:SQMLimits}
\begin{figure}
\begin{center}
\resizebox{8.0cm}{7.5cm}{\includegraphics{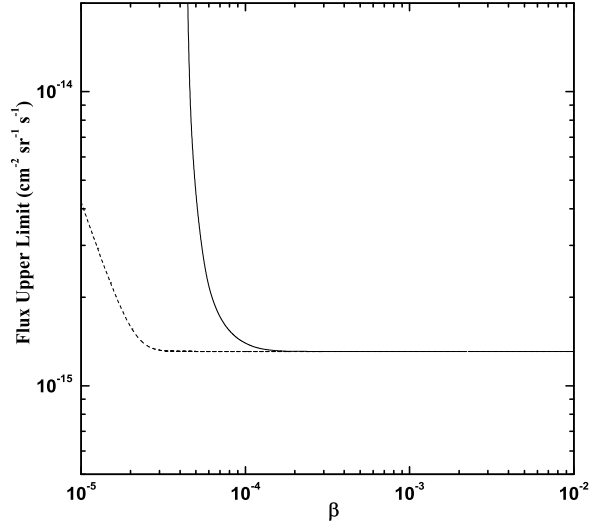}}
\caption{\small 90\% C.L. flux upper limits versus $\beta$ for a downgoing flux of nuclearites obtained by the SLIM experiment at Chacaltaya. The solid line is valid for nuclearites with $M_{N} < 8.4\cdot 10^{14}$ GeV/c$^{2}$, the dashed line is for $M_{N}\sim10^{17}$ GeV/c$^{2}$.}
\label{fig:NuclUpLimit}
\end{center}
\end{figure}

\begin{figure}
\begin{center}
\resizebox{8.0cm}{7.5cm}{\includegraphics{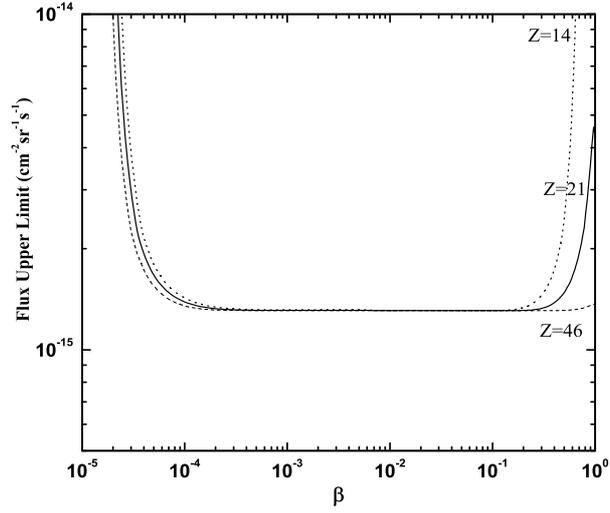}}
\caption{\small 90\% C.L. upper limits versus $\beta$ for a flux of light downgoing strangelets obtained by SLIM at Chacaltaya. The dotted line is for a charge $Z =14$ at the detector, the solid line is for $Z =21$ and the dashed line for $Z=46$, respectively.}
\label{fig:StrUpLimit}
\end{center}
\end{figure}

The analysis of 427 m$^{2}$ of Nuclear Track Detectors exposed at the Chacaltaya high altitude Lab for  4.22 years yielded no candidate events. This null observation allows us to set upper limits on the fluxes of nuclearites and strangelets. For downgoing nuclearites with $\beta\sim10^{-3}$ we obtain a 90\% C.L. flux upper limit of $\sim$$1.3\cdot 10^{-15}$ cm$^{-2}$ s$^{-1}$ sr$^{-1}$, as shown in Fig.~\ref{fig:NuclUpLimit}. This upper limit is also valid for downgoing strangelets with velocities at the detector level, in the $\beta$ range $2\cdot 10^{-4} \div 10^{-1}$ as indicated in Fig.~\ref{fig:StrUpLimit}. These velocities are dependent on the propagation model and on the atmospheric depth crossed by the initial strangelet.

\section{Q-balls}
\label{sec:Qballs}

Q-balls are non\-topological solitons predicted by minimal supersymmetric generalizations of the Standard Model of particle physics. They are hypothesized supersymmetric coherent states of squarks, sleptons and Higgs fields, and may carry some conserved global baryonic charge Q.

According to their properties of interaction with matter~\cite{Kusenko,Ouchrif}, Q-balls are classified into two groups : $(i)$ neutral Q-balls, generally called SENS (Supersymmetric Electrically Neutral Solitons), that should be massive and may catalyse proton decay; and, $(ii)$ charged Q-balls called SECS (Supersymmetric Electrically Charged Solitons), that might be formed by the SENS gaining an integer electric charge from proton or nuclei absorption. Coulomb barrier may prevent further absorption of nuclei.

	In refs.~\cite{Kusenko,Kusenko05} it was pointed out that neutral Q-balls (SENS) would interact with matter mainly by converting nuclei into pions with a cross section determined from their size. This process occurring at a higher rate than ordinary collisions of Q-balls with nuclei, may not be relevant for detection in NTDs. However, it could lead to SENS acquiring an electric charge, thus transforming into SECS. SECS could capture electrons and the resulting system would be similar to an atom with an enormous heavy nucleus~\cite{Bakari}. The cross section for interaction with matter will be determined by the size of the electronic cloud. Thus, charged Q-balls should interact with matter in a way not too different from nuclearites, i.e. their energy losses would be about the same as for nuclearites with a constant radius $\sim$$10^{-8}$ cm (given by the size of the electronic cloud). The conclusions  obtained for nuclearites can be extended to SECS of medium-high masses. For SECS with very large masses, $M_{Q} > 10^{26}$ GeV/$c^{2}$ or $M_{Q} > 10^{30}$ GeV/$c^{2}$ (depending on the energy scale of the SUSY breaking theory) the core radius will be larger than $10^{-8}$ cm. Consequently, the electrons are inside the Q-ball and the whole system would behave as SENS~\cite{Arafune}.

\begin{figure}
\begin{center}
\resizebox{8.5cm}{8.5cm}{\includegraphics{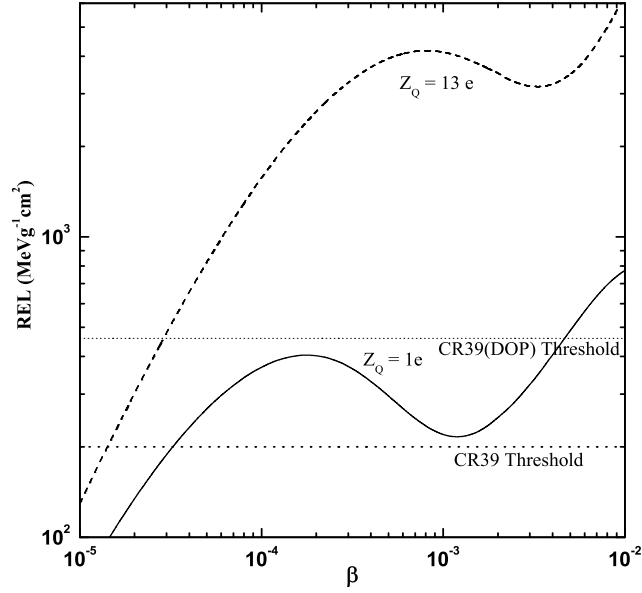}}
\caption{\small Restricted Energy Losses of charged Q-balls with electric charge $Z_{Q} =1$e and $13$e, as a function of their velocity in CR39 nuclear track detectors. The detector threshold is also shown (dotted horizontal lines).}
\label{fig:QballREL}
\end{center}
\end{figure}

	In the case where SECS would not capture electrons, a different method is needed to estimate the rate of energy losses~\cite{Arafune}. For typical galactic velocities $v\sim10^{-3}c$, there are two contributions to their energy loss which come from the interaction with electrons and nuclei in matter~\cite{Bakari,Kusenko,Ouchrif}. Electronic energy losses are dominant at the highest energies and can be computed from a Ziegler fit to the experimental data~\cite{SRIM}. Energy losses from nuclear collisions are computed from the diffusion cross sections considering a screened electromagnetic potential as described in~\cite{Ouchrif} and references therein. The REL for Q-balls of two different charges $Z_{Q} =1$e and $13$e is illustrated in Fig.~\ref{fig:QballREL}. The CR39 and CR39(DOP) thresholds are also shown. It can be seen that these SECS, for velocities $\beta\ge3.5\cdot10^{-5}$, could be detected in SLIM CR39. However, in the case of CR39(DOP) and for SECS with charge $Z_{Q} =1$e, the range of detectability is reduced to $\beta>4\cdot10^{-3}$.

\begin{figure}
\begin{center}
\resizebox{8.5cm}{8.5cm}{\includegraphics{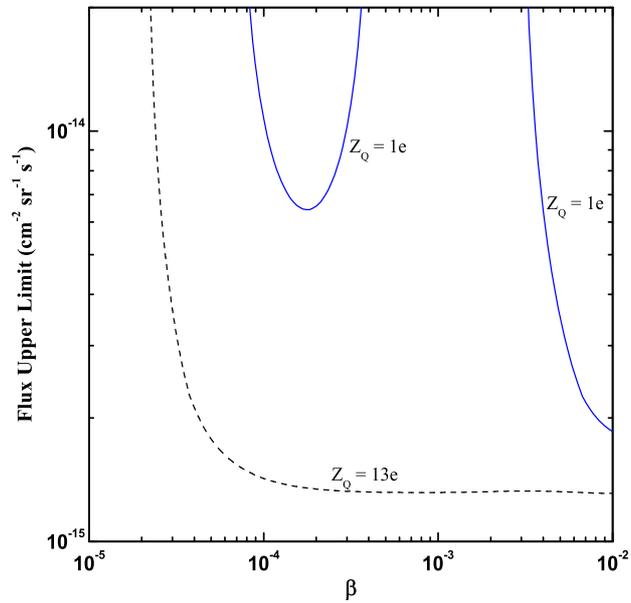}}
\caption{\small 90\% C.L. upper limits obtained for a downgoing flux of charged Q-balls (SECS) with $Z_{Q} =1$e and $Z_{Q} =13$e, plotted vs $\beta$.}
\label{fig:QballUpLimit}
\end{center}
\end{figure}

	Since no candidate event was found in the SLIM experiment, we set the flux upper limits as shown in Fig.~\ref{fig:QballUpLimit}, for Q-balls with $Z_{Q}\ge1e$. It can be seen that, for Q-balls with $Z_{Q}\sim13e$ and $\beta>10^{-4}$, the limit is about $1.3\cdot 10^{-15}$ cm$^{-2}$s$^{-1}$sr$^{-1}$. While, for Q-balls with very low charges it is considerably higher.

\section{Conclusions}
\label{sec:conclusion}
The analysis of the CR39 NTDs of the SLIM experiment has resulted in new upper limits (90\% C.L.) on the flux of SQM nuggets and charged Q-balls at the level of $1.3\cdot 10^{-15}$ cm$^{-2}$s$^{-1}$sr$^{-1}$. 

\begin{figure}
\begin{center}
\resizebox{8.2cm}{8.1cm}{\includegraphics{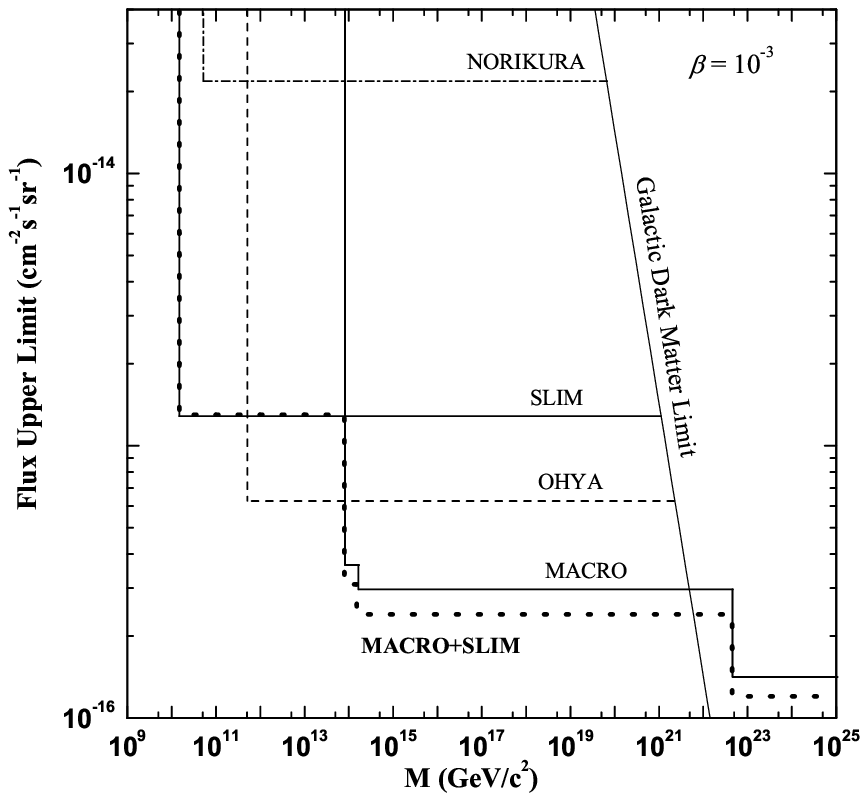}}
\caption{\small Flux upper limits vs mass, for intermediate and high mass nuclearites (with $\beta\sim10^{-3}$), given by various searches with NTDs, see text. The combined flux from MACRO and SLIM is also shown.}
\label{fig:NuclLimitComp}
\end{center}
\end{figure}

	In Fig.~\ref{fig:NuclLimitComp} the flux upper limits for nuclearites with $\beta>10^{-3}$ set by other searches based on NTDs \cite{Macro,Ohya,Norikura}, together with the SLIM limit and the combined MACRO + SLIM limits are shown. The galactic dark matter upper bound is also indicated. Above $4\cdot10^{22}$ GeV/$c^{2}$ the upper limit for a $4\pi$ acceptance is also drawn.

\begin{figure}
\begin{center}
\resizebox{8.2cm}{8.2cm}{\includegraphics{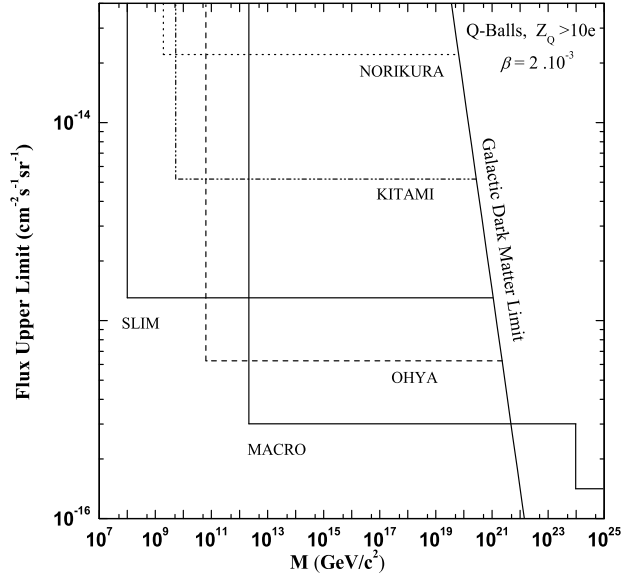}}
\caption{\small Flux upper limits of charged Q-balls with $Z_{Q} >10$e versus mass, see text.}
\label{fig:QballLimitComp}
\end{center}
\end{figure}

\begin{figure}
\begin{center}
\resizebox{8.8cm}{8.8cm}{\includegraphics{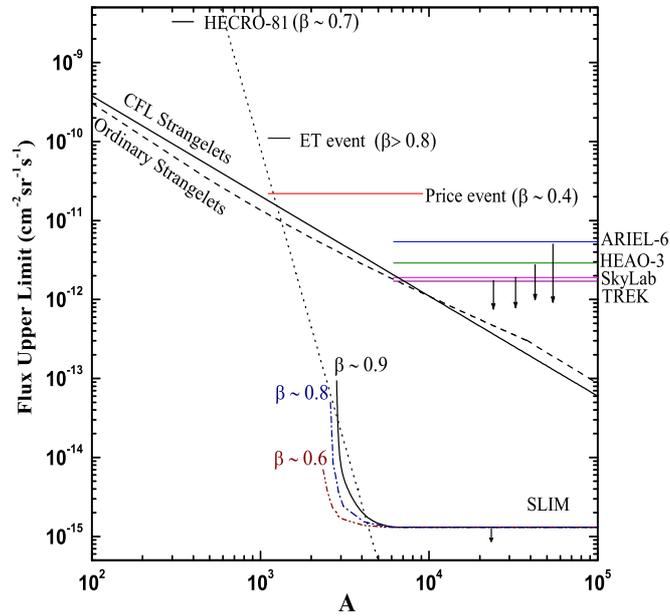}}
\caption{\small Flux upper limits vs A for relativistic strangelets as determined by experiments onboard balloon and space and by SLIM at mountain altitude (see text). The lines are the expected fluxes from different models: Madsen~\cite{Madsen05} solid (CFL Strangelets) and dashed (ordinary strangelets) lines; Wilk~\cite{WilkF} dotted line. The SLIM upper limit is shown for three different velocities ($\beta$) at the detector level.}
\label{fig:StrLimitComp}
\end{center}
\end{figure}

	Fig.~\ref{fig:QballLimitComp} shows the flux upper limits for charged Q-balls (for $Z_{Q}>10$ e and $\beta$$\sim$$2\cdot10^{-3}$) from various experiments employing NTDs (\cite{Macro} and \cite{Ohya,Norikura,Kitami} as re-estimated in ref.\cite{Arafune}) together with the SLIM result. 

	Finally, the flux upper limits set by SLIM for relativistic strangelets are compared with the results from previous searches with experiments onboard stratospheric balloons and in space (ARIEL-6 \cite{Ariel}, HEAO-3 \cite{Heao}, SkyLab \cite{Skylab} and TREK \cite{Trek}) as shown in Fig.~\ref{fig:StrLimitComp}. The candidate ``exotic'' events reported so far are also shown \cite{Hecro,ET,Price}.

	In order to compare searches performed by SLIM with other experiments we had to assume a particular propagation model in atmosphere.

	According to Wilk \cite{WilkF}, the number of events predicted to reach Chacaltaya and be detected with SLIM is $\sim$$6$ events/100 m$^{2}$/year/sr. This would correspond to $\sim$$10$ events with $Z\ge14$ and velocities $<0.7$c at the detector level. For higher velocities, the flux would be $\sim$$1$ event/100 m$^{2}$/year/sr for strangelets with masses $A>2800$ at the top of the atmosphere. Taking into account SLIM angular acceptance~\cite{Balestra0801}, the number of events that could have been recorded by SLIM NTDs for strangelet charge $Z$$\sim$$18$ at the detector level is $\sim 5$ events.
	
	Assuming the Banerjee propagation model \cite{Banerjee,BanerjeeF}, the number of events with charge $Z>14$ that should have been detected at Chacaltaya is $\sim$$20\div40$ events/year. As our detector acceptance is larger at the expected velocities ($<0.7c$) and the penetration through the atmosphere has no suppression effect in the Banerjee model, SLIM should have recorded about $80$ events. The null observation tends to exclude the Banerjee propagation model of strangelets in the atmosphere, while the Wilk model is only marginally excluded. 
	
	At the moment we do not discuss the predictions made in ref.~\cite{Madsen05} since the comparison may not be straightforward.
	
	In conclusion, SLIM has extended the search for intermediate mass exotic particles with observations at high altitude and lowered, in combination with MACRO, the flux upper limit for larger mass exotics. New results on the strangelet flux at lower masses ($A<1000$) may hopefully come from next generation satellite experiments e.g. AMS-02 \cite{Finch}, PAMELA \cite{Pamela}, ...

\section*{Acknowledgments}

We acknowledge the collaboration of S. Balestra, E. Bottazzi, L. Degli Esposti and G. Grandi of INFN Bologna and the technical staff of the Chacaltaya Laboratory. We thank INFN and ICTP for providing grants for non-italian citizens.

\end{document}